\documentclass{aa}  
\usepackage{graphicx}
\usepackage{txfonts}
\usepackage{latexsym}
\usepackage{natbib}
\usepackage{comment}
\usepackage{subfigure}
\usepackage{subfig}
\usepackage{inputenc}
\usepackage{float}
\begin{document}

\title{Investigating the nature of the extended structure around the Herbig star RCrA using integral field and high-resolution spectroscopy}


\author{E.\,Rigliaco\inst{1}
\and R.\,Gratton\inst{1} 
\and D.\,Mesa\inst{1}
\and V.\,D'Orazi\inst{1}
\and M.\,Bonnefoy\inst{2}
\and J.M.\,Alcal\`a\inst{3}
\and S.\,Antoniucci\inst{4}
\and F.\,Bacciotti\inst{5}
\and M.\,Dima\inst{1}
\and B.\,Nisini\inst{4}
\and L.\,Podio\inst{5}
\and M.\,Barbieri\inst{6}
\and R.\,Claudi\inst{1}
\and S.\,Desidera\inst{1}
\and A.\,Garufi\inst{5}
\and E.\,Hugot\inst{7}
\and M.\,Janson\inst{8,9}
\and M.\,Langlois\inst{10}
\and E.L.\,Rickman\inst{11}
\and E.\,Sissa\inst{1} 
\and M.\,Ubeira Gabellini\inst{12}
\and G.\,van der Plas\inst{2}
\and A.\,Zurlo\inst{13,14,8}
\and Y.\,Magnard,\inst{15}
\and D.\,Perret\inst{16}
\and R.\,Roelfsema\inst{17}
\and L.\,Weber\inst{18}
}
\institute{
\inst{1}INAF/Osservatorio Astronomico di Padova, Vicolo dell'osservatorio 5, 35122 Padova
\email{elisabetta.rigliaco@inaf.it}\\
\inst{2}Univ. Grenoble Alpes, CNRS, IPAG, 38000 Grenoble, France\\
\inst{3}INAF-Osservatorio Astronomico di Capodimonte, Salita Moiariello 16, 80131 Napoli, Italy \\
\inst{4}INAF-Osservatorio Astronomico di Roma, via di Frascati 33, 00078 Monte Porzio Catone, Italy \\
\inst{5}INAF-Osservatorio Astrofisico di Arcetri, Largo E. Fermi 5, I-50125, Firenze, Italy \\
\inst{6}INCT, Universidad De Atacama, calle Copayapu 485, Copiap\'{o}, Atacama, Chile\\
\inst{7}Aix Marseille Univ, CNRS, CNES LAM, Marseille, France \\
\inst{8}Max-Planck-Institut f\"ur Astronomie, K\"onigstuhl 17, 69117, Heidelberg, Germany \\
\inst{9}AlbaNova University Center, Stockholm University, Stockholm, Sweden \\
\inst{10}Univ. Lyon, Univ. Lyon 1, ENS de Lyon, CNRS, CRAL UMR 5574, 69230 Saint-Genis-Laval, France \\
\inst{11}Geneva Observatory, University of Geneva, Chemin des Maillettes\\
\inst{12}Dipartimento di Fisica, Universit\'a Degli Studi di Milano, Via Celoria, 16, I-20133 Milano, Italy \\
\inst{13}Nucleo de Astronomia, Facultad de Ingenieria y Ciencias, Universidad Diego Portales, Av. Ejercito 441, Santiago, Chile \\
\inst{14}Escuela de Ingenieria Industrial, Facultad de Ingenieria y Ciencias, Universidad Diego Portales, Av. Ejercito 441, Santiago, Chile \\
\inst{15}Univ. Grenoble Alpes, CNRS, IPAG, F-38000 Grenoble, France.\\ 
\inst{16}LESIA, Observatoire de Paris, Universit\'e PSL, CNRS, Sorbonne Universit\'e, Univ. Paris Diderot, Sorbonne Paris Cit\'e, 5 place Jules Janssen, 92195 Meudon, France\\ 
\inst{17}NOVA Optical Infrared Instrumentation Group, Oude Hoogeveensedijk 4, 7991 PD Dwingeloo, The Netherlands\\
\inst{18}Geneva Observatory, University of Geneva, Chemin des Mailettes 51, 1290 Versoix, Switzerland\\
} 

   \date{Received : 16 September 2019; accepted: 30 September 2019 }

 
  \abstract
 {We present a detailed analysis of the extended structure detected around the young and close-by Herbig Ae/Be star R~CrA. This is a young triple system with an intermediate mass central binary whose separation is of the order of a few tens of the radii of the individual components, and an M-star companion at about 30 au.}
 {Our aim is to understand the nature of the extended structure by means of combining integral-field and high-resolution spectroscopy. }
 {We conducted the analysis based on FEROS archival optical spectroscopy data and adaptive optics images and integral-field spectra obtained with SINFONI and SPHERE at the VLT. }
 {The observations reveal a complex extended structure that is composed of at least two components: a non-uniform wide cavity whose walls are detected in continuum emission up to 400~au, and a collimated wiggling-jet detected in the emission lines of Helium and Hydrogen. Moreover, the presence of [Fe~{\sc{ii}}] emission projected close to the cavity walls suggests the presence of a slower moving wind, most likely a disk wind. The multiple components of the optical forbidden lines also indicate the presence of a high-velocity jet co-existing with a slow wind. We constructed a geometrical model of the collimated jet flowing within the cavity using intensity and velocity maps, finding that its wiggling is consistent with the orbital period of the central binary. The cavity and the jet do not share the same position angle, suggesting that the jet is itself experiencing a precession motion possibly due to the wide M-dwarf companion. }
 {We propose a scenario that closely agrees with the general expectation of a magneto-centrifugal-launched jet. These results build upon the extensive studies already conducted on R~CrA. }


\titlerunning{Investigating the nature of the extended structure around the Herbig star RCrA}
\authorrunning{Rigliaco et al.}
  \maketitle
%

\section{Introduction}

Herbig Ae/Be stars \citep{Herbig1960} are pre-main sequence stars of intermediate mass covering the range between low-mass T Tauri stars (TTSs), and the embedded massive young stellar objects. They are considered the high-mass counterparts of pre-main sequence T Tauri stars  \citep{Strom1972, Cohen1979, Finkenzeller1984}. These stars, like T Tauri stars, show rich emission-lines spectra, infrared continuum excess and veiled photospheric absorption. The formation of stars in the low and intermediate-mass regimes involves accretion disks, and fast collimated outflows and jets.  The accretion activity is established from a spectroscopic point of view through the presence of emission lines in the stellar spectrum, in wavelength ranges that span from the ultraviolet to the infrared (e.g. \citealt{Alcala2017, Mendigutia2012}). Jets and outflows in TTSs are also spectroscopically revealed by the analysis of emission lines in their spectrum (e.g., \citealt{Edwards2007, Nisini2018}), while there is instead a paucity of detected jets and outflows around intermediate mass Herbig Ae/Be stars (e.g., \citealt{Grady2003, Ellerbroek2014}). This is partially due to the shorter time the intermediate mass objects spend in their pre-main sequence phase. 

The advent of high-contrast scattered light observations of protoplanetary disks around TTSs and Herbig Ae/Be stars performed with Adaptive Optics techniques using instruments such as the Gemini Planet Imager (GPI: \citealp{Macintosh2014}) or the Spectro-Polarimetric High-contrast Exoplanet REsearch (SPHERE: \citealp{Beuzit2019}) has allowed for the investigation of the immediate surroundings of these stars, revealing a wealth of extended structures: concentric rings (e.g. \citealp{Ginski2016, Perrot2016, Feldt2017}), cavities (e.g. \citealp{Pohl2017, Ligi2018}), spiral arms (e.g. \citealp{Maire2017, Benisty2017}), and asymmetries. ALMA, on the other hand, has allowed for the investigation of the immediate surroundings of younger and more embedded stars, revealing in turn non-axisymmetric features (e.g. \citealp{van2013major}), multiple narrow rings (e.g. \citealp{Perez2016, fedele2017, fedele2018, Huang2018}). These features are usually associated with circumstellar/protoplanetary disks, but there are a few cases where the images reveal the presence of extended and elongated structures in the jet directions. The analysis of these elongated structures from Herbig stars is important because it allows us to study if the origins of the observed features are similar to T Tauri stars, that is, linked to interaction between magnetic fields and the circumstellar environment \citep{Hubrig2018}.
Only a few observations of jets around close-by (less than 200~pc from the Sun) Herbig stars exist. A collimated jet around the Herbig Ae star HD~163296 was detected in Ly$\alpha$ by \citet{Devine2000} and \citet{Grady2000}. A bipolar jet is also driven by the Herbig Ae stars MWC480 and HD104237 as shown by \cite{Grady2010, Grady2004}. Recently, \cite{Garufi2019} have revealed, in scattered light, a dip along the jet axis around the intermediate mass star RY~Tau that is consistent with an outflow cavity carved in the ambient envelope by the jet and the wind outflow. Further out Herbig stars have also been observed driving jets: extended emissions from both components of the Z~CMa system have been revealed by \cite{Antoniucci2016} using the ZIMPOL instrument (Zurich Imaging Polarimeter) of SPHERE at the Very Large Telescope (VLT). 
A jet and counter-jet were also revealed in LkHa233 using the Hubble Space Telescope data \citep{Melnikov2008}. The paucity of observations of jets around Herbig stars with respect to the higher number of jets observed around T Tauri stars might be either due to an intrinsic abundance of TTSs, or to a shorter timescale of these structures around Herbig stars. In either case, any new observation of jets around Herbig stars is essential to shedding light on the accretion/ejection mechanisms operating on Herbig stars. 

R~CrA (HIP~93449) is an ideal target to enlarge the number of Herbig Ae/Be stars where jet-like structures are observed. It is the brightest member of the Coronet Cluster, belonging to the Corona Australis star-forming region, which is one of the nearest and most active regions of ongoing star formation. The Coronet Cluster is highly obscured \citep{Taylor1984}, and characterized by high and spatially-variable extinction with A$_V$ up to 45 mag \citep{Neuhauser2008}. 
This star has been extensively studied over the years. 
\citet{Takami2003} suggested the existence of a companion and of an outflow to explain the positional photo-center displacement observed in spectro-astrometric observations both in the blue and red-shifted wing of the spectrally resolved H$\alpha$-line. The spectral type of R~CrA has also been largely debated. In analyzing IRAS data, \citet{Bibo1992} found  spectral type B8, stellar radius of 3.1~R$_{\odot}$, bolometric luminosity between 99-166~L$_\odot,$ and mass 3.0~M$_{\odot}$. Spectral type A5 and L$_{bol}$=92~L$_\odot$ were found by \citet{Chen1997}, while R~CrA was classified as F5 by \citet{Hillenbrand1992, Natta1993, GarciaLopez2006}. R~CrA is in a particularly early evolutionary phase \citep{Malfait1998}, because it is still embedded in its dust envelope, whose emission dominates the spectral energy distribution from mid-IR to millimeter wavelengths \citep{Kraus2009}. At optical wavelengths, the star is known to be highly variable, both on long and short time scales \citep{Bellingham1980}. \citet{Sissa2019} found  A$_V$=5.47$\pm$0.5~mag, slightly larger than the value obtained by \citet{Bibo1992} (A$_V$=4.65~mag). R~CrA shows indications of active accretion, and various outflow tracers have been reported. For instance, a compact bipolar molecular outflow with an east-west orientation \citep{Walker1984, Levreault1988, Graham1993}, as well as several Herbig-Haro objects (in particular HH 104 A/B), have been associated with R~CrA \citep{Hartigan1987, Graham1993}. However, more recent studies \citep{Anderson1997, Wang2004} convincingly identified the source IRS 7 as the driving source of these outflows, making a physical association with R~CrA rather unlikely. 
The outflow was also detected in infra-red (IR) through CO emission (at 4.6$\mu$m) using astrospectrometry by \citet{vanderPlas2015}. They notice that the CO emission is located entirely on one side of the IR continuum emission, and blueshifted by $\sim$10~km/s with respect to the surrounding Corona Australis molecular cloud. They point out that CO emission in the R CrA spectrum is likely due to an outflow, as also suggested from the bow-shocks seen in shocked H2 emission in the immediate vicinity of R CrA (\citealt{Kumar2011}). 
The study of R~CrA has gained new momentum in recent years, thanks to  high-contrast imaging observations obtained with SPHERE \citep{Mesa2019} and the NAOS-CONICA at VLT \citep{Cugno2019}. These studies focused their attention on the detection of a stellar companion as close as 19-28 ~au, and with a mass range between 0.1-1.0~M$_{\odot}$. Moreover, in the \cite{Mesa2019} analysis, the presence of an elongated jet-like structure was pointed out, together with some evidence of a disk seen almost edge-on. Recently, \citet{Sissa2019} analyzed the light curve of the star, and found that the central object of R~CrA is a binary with masses of 3.0 and 2.3~M$_{\odot}$ for the two components and with a period of approximately 66 days. Together with the discovery of the M-type companion, it makes it a triple system. 

In this paper, we analyze images of R~CrA acquired with SPHERE and the SINFONI integral field spectrograph at the VLT, and an archival optical spectrum from the Fibre-fed Optical Echelle Spectrograph (FEROS) archive. In Sect.~2, we describe the  collected data. in Sect.~3, we describe the data analysis. In Sect.~4, we propose a scenario that reconciles all the findings, and in Sect.~5, we summarize and conclude.

\section{Observations and data reduction} 

As mentioned in the previous section, several studies at different wavelength ranges have been conducted over the years on R~CrA. The use of high-contrast imagers has made it possible to shed new light on this interesting source. In this section, we describe the observations of R~CrA acquired in recent years with SPHERE and SINFONI at VLT  (summarized in Table~\ref{log_tab}), and FEROS at the 2.2~m telescope in La Silla. We focus here on the elongated structure already reported in \citet{Mesa2019}, and identified in Figure~\ref{IFS_total}, analyzing the emission spectrum of the source, and investigating the origin of the emitting lines in light of images of the jet-like structure observed in the SPHERE data.

\begin{table*}
\caption{Log of the observations obtained with SPHERE and SINFONI}
\label{log_tab} 
\centering            
\begin{tabular}{c c c c c c c}
\hline\hline
Date  & Instrument & band &  FOV & Spaxel & Spectral resolution & Wavelength coverage\\
\hline      
2018-06-19 & IFS &  Y-H & 1.7$^{\prime\prime} \times$1.7$^{\prime\prime}$ & 7.49mas$^2$ & $\sim$30 & 0.95--1.65$\mu$m \\
2018-06-19 & IRDIS & K1, K2 & 11.0$^{\prime\prime} \times$11.0$^{\prime\prime}$ & 12.25mas$^2$ & --- & 2.11$\mu$m,2.25$\mu$m \\
2018-09-11 & SINFONI & H & 0.8$^{\prime\prime} \times$0.8$^{\prime\prime}$ & 12.5mas$^2$ & $\sim$3000 & 1.45--1.85$\mu$m \\
\hline      
\end{tabular}
\end{table*}

\subsection{SPHERE data}

\begin{figure}
\begin{subfigure}
  \centering
  \includegraphics[width=\linewidth]{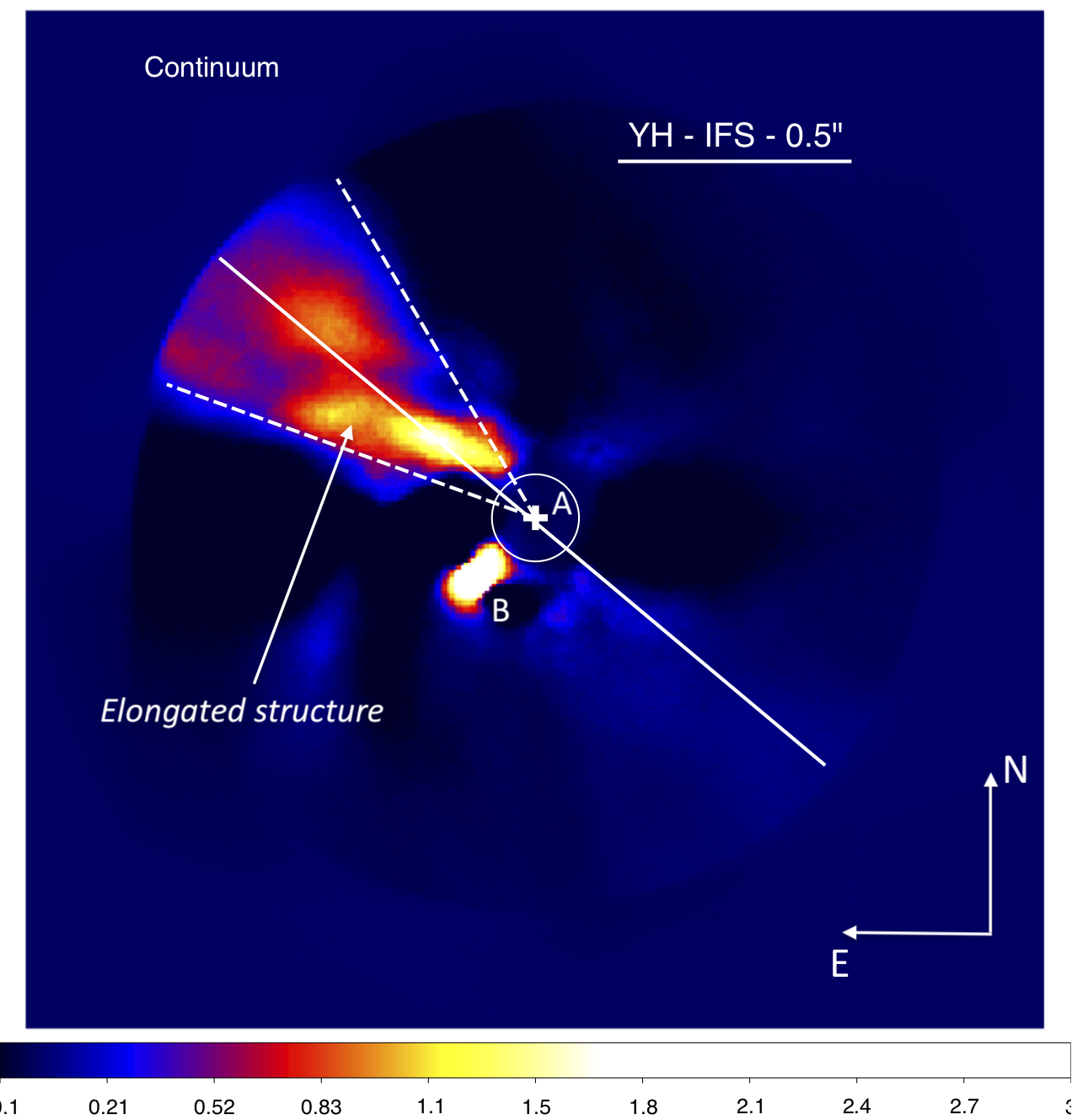}  
\end{subfigure}
\begin{subfigure}
  \centering
  \includegraphics[width=\linewidth]{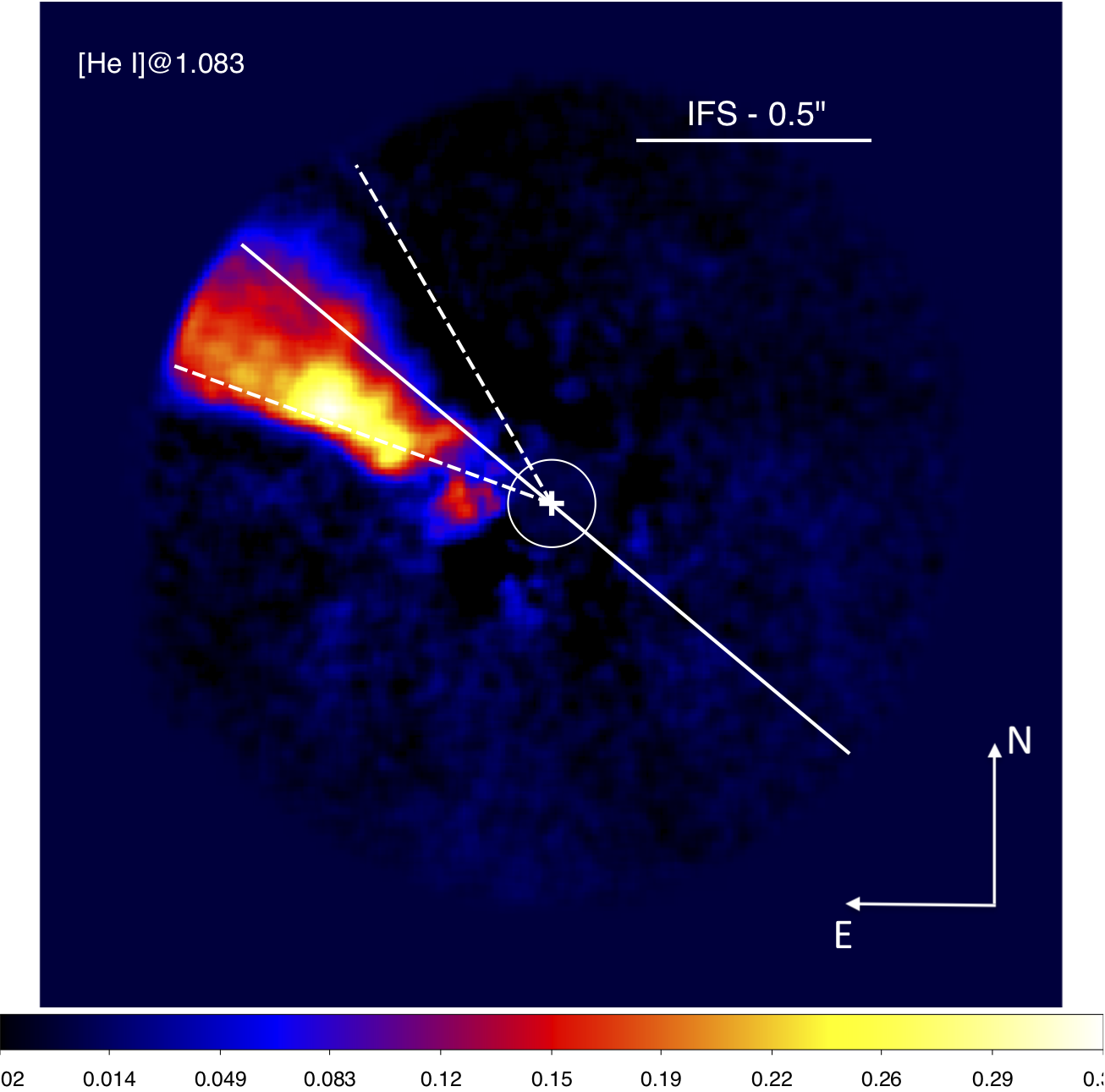}  
\end{subfigure}
\caption{Top panel: Median image obtained using monochromatic PCA with 1 principal component for spectral channel on SPHERE-IFS image, showing the continuum emission. The bright source at about 0.3~arcsec south-east from the center, identified with B, is the M-star companion detected by \citet{Mesa2019}, which appears elongated as a consequence of the differential imaging adopted. Solid and dashed lines refer to the median PA and aperture of the "elongated structure" as measured in Sect. 3.2, respectively.  Bottom panel:  Same image as top panel, but only in channels containing Helium I at 1.083~$\mu$m line. The white solid and dashed lines show the PA of the extended structure, as discussed in Sect.~3. The white circle shows the size of the coronagraph used. }
\label{IFS_total}
\end{figure}

R~CrA was observed with SPHERE \citep{Beuzit2019} in four different epochs, in coronagraphic mode, as described in \cite{Mesa2019}. In this paper, we use the observation of the night 2018-06-19 as part of the SHINE (SpHere INfrared survey of Exoplanets - \citealt{Chauvin2017}) guaranteed time observations, since they were taken in better weather conditions and reached deeper contrast with respect to the other sets of observations. The instrument was used in  IRDIFS$\_$EXT mode, allowing simultaneous observation with the integral-field spectrometer IFS \citep{Claudi2008} and with the dual band differential imager and spectrograph IRDIS \citep{Dohlen2008}. In this mode, IFS provides diffraction-limited observations covering the Y and H bands (0.95-1.65~$\mu$m), with a spectral resolution of $\sim$30 inside the 1.7$^{\prime\prime} \times$ 1.7$^{\prime\prime}$ field of view. The IRDIS sub-system, also diffraction-limited, was set in its dual-band imaging mode (DBI, \citealt{Vigan2010}) to simultaneously observe with the K1 and K2 filters (K1=2.110~$\mu$m and K2=2.251~$\mu$m, width 0.1~$\mu$m) with a 11.0$^{\prime\prime} \times$ 11.0$^{\prime\prime}$ field of view. Details on the observing mode and data reduction are reported in \citet{Mesa2019}. Data were acquired in pupil stabilized mode, with a sequence of acquisition, while the field of view (FOV) rotated. This allows application of angular differential imaging (ADI, \citealp{Marois2006}) to reduce the impact of speckle noise: in general, we use an approach based on a principal component analysis (PCA, \citealp{Soummer2012}) for differential imaging. The IFS data allows a further spectral dimension of the hyper-data cube. The IFS image, obtained using a PCA done  along temporal and spectral channels  (ASDI) simultaneously, with 10 principal components (\citealt{Mesa2015}) by \cite{Mesa2019}, highlighted the presence of an elongated structure in the north-east direction, as well as the hint of a disk seen almost edge-on in the image. In the following we refer to this  "elongated structure", deferring its interpretation to the next sections. 

\begin{figure}
\begin{subfigure}
  \centering
  \includegraphics[width=\linewidth]{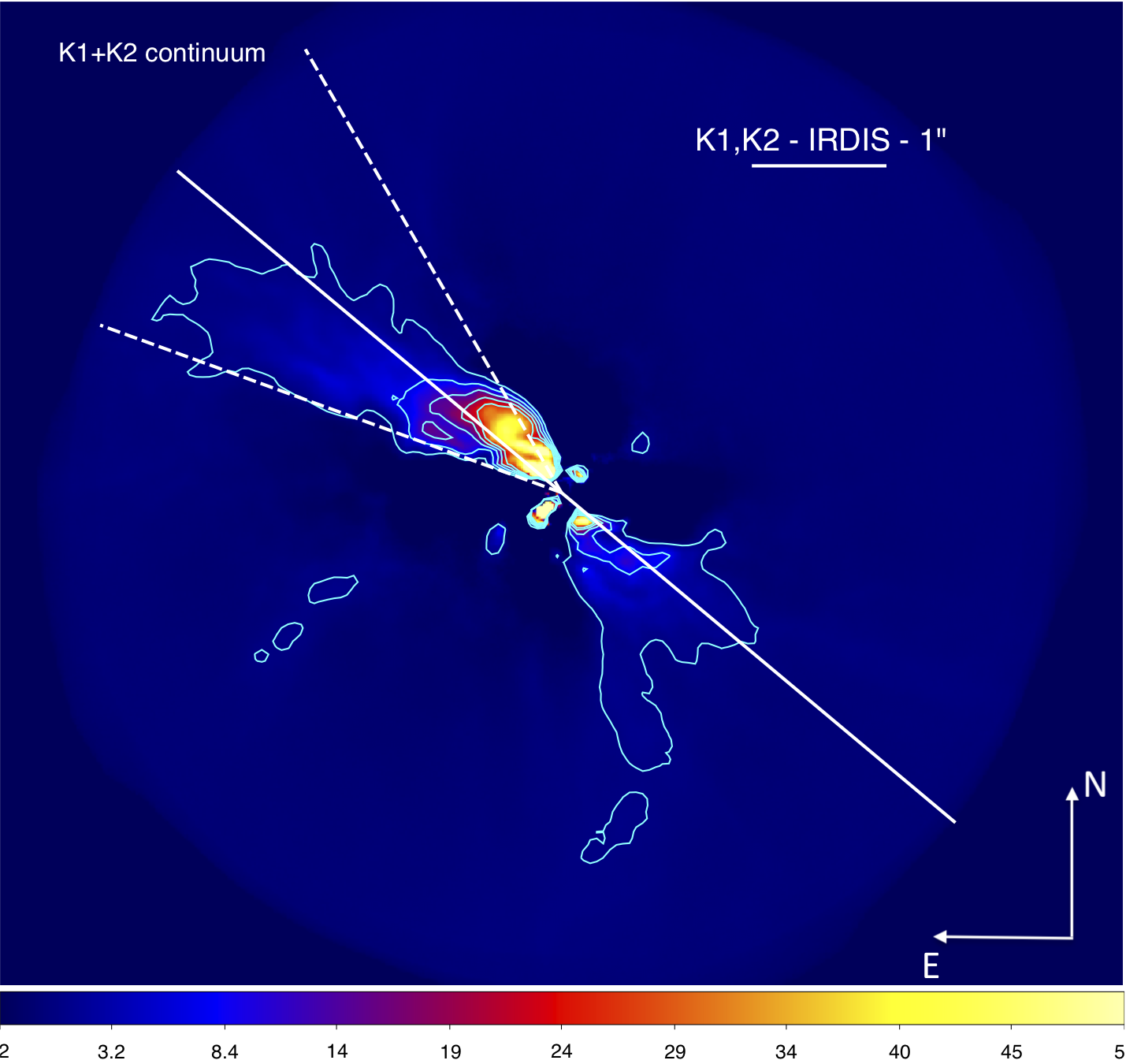}  
\end{subfigure}
\begin{subfigure}
  \centering
  \includegraphics[width=\linewidth]{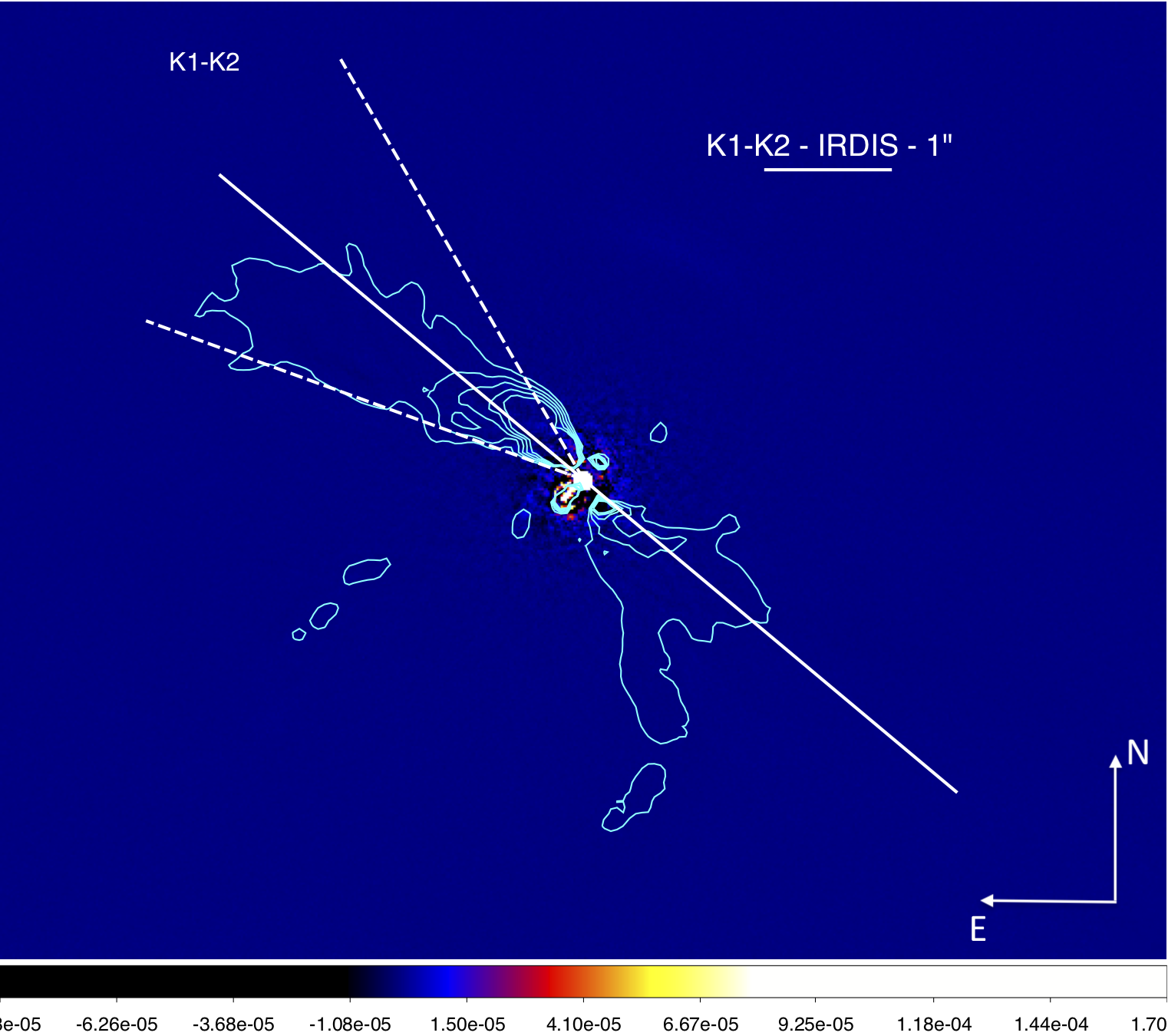}  
\end{subfigure}
\caption{Top panel: Median image obtained using monochromatic PCA with one principal component per spectral channel on SPHERE-IRDIS image, showing continuum emission. Contours are also shown. Bottom panel:  Subtraction of K1-K2 filters. No signal is left. For comparison, the contours used in the top panel are also reported. Solid and dashed lines as in Figure~\ref{IFS_total}.}
\label{IRDIS_total}
\end{figure}

In the present paper, we prefer to use images obtained with a PCA done separately on the different spectral channels (monochromatic PCA) in order to retrieve all the information coming from every channel; this makes it possible to highlight the different wavelengths sampled in the image. In addition, we also used images that are obtained simply subtracting a radial-profile and that are not affected by the self-subtraction of non-uniform imaged structures typical of differential imaging. 
The median IFS image showing the continuum emission in Y and H band and the most prominent line emission of the Helium at 1.083$\mu$m is shown in Figure~\ref{IFS_total}. 
Figure~\ref{IRDIS_total} shows the continuum emission obtained in K1 and K2 with IRDIS, and the subtraction of the K1-K2 filters. 

\subsection{SINFONI data} 
 
\begin{figure*}[!h]
\centering
\includegraphics[angle=0,width=\hsize]{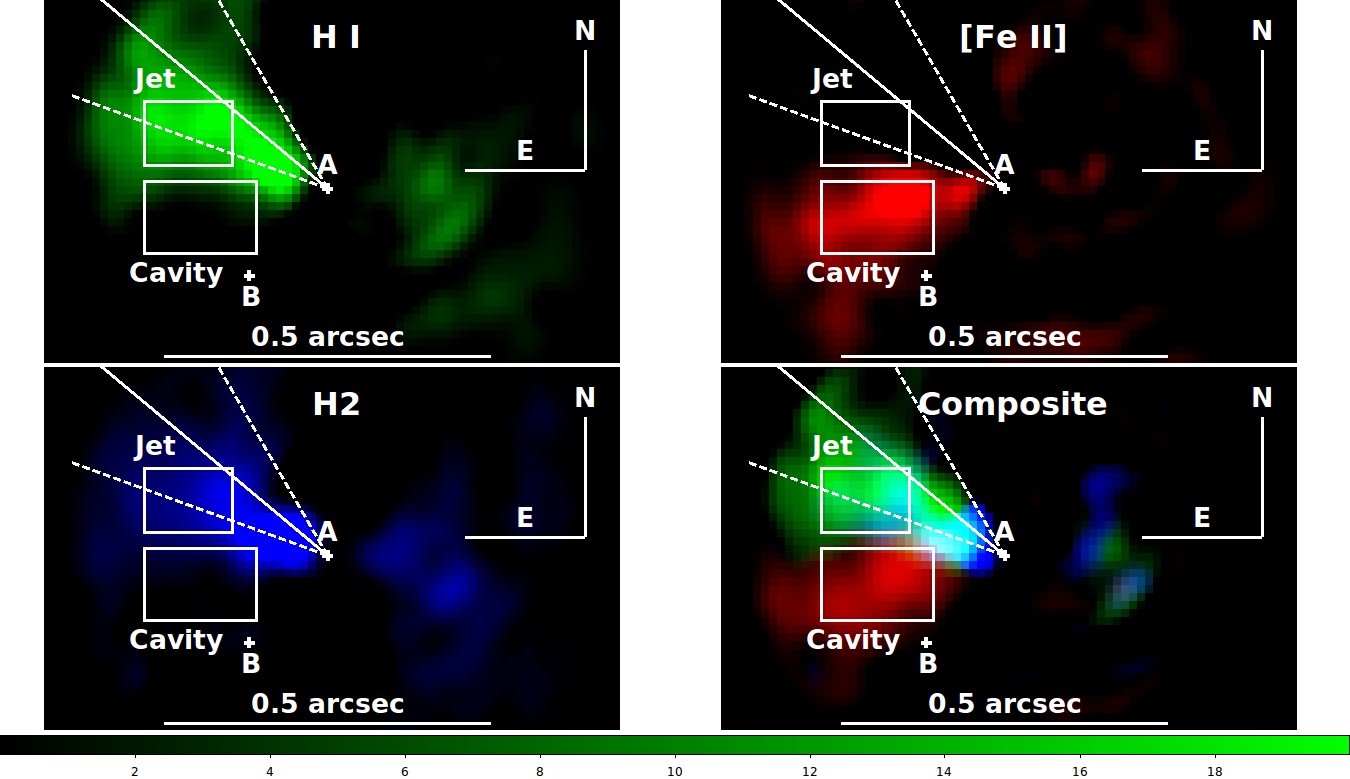}
\caption{False-color images obtained from SINFONI data cube in line emission. In green, the summed emission from the Hydrogen lines, in blue, the sum of the emission from the H$_2$ lines, and in red, the sum of the emission from the [Fe~{\sc{ii}}] lines. Bottom-right panel shows the composite image obtained from the sum of the other three panels. The boxes show the regions where we extracted the spectra for the "jet" and "cavity", shown in Figure~\ref{Fig:spectra_sinfoni}. 
Solid and dashed lines as in Figure~\ref{IFS_total}: they indicate to the "elongated structure" PA and aperture as identified from IFS and IRDIS continuum images.}
\label{Fig:composite_2}
\end{figure*}

A H-band image of R~CrA was obtained with the AO-fed integral field spectrograph SINFONI \citep{Eisenhauer2003, Bonnet2004} operating between 1.45--1.85~$\mu$m with a resolution R$\sim$3000. The data were collected during the night 2018-09-11 under the program 2101.C-5048(A) (P.I. D.Mesa) with a spatial sampling of 0.0125$^{\prime\prime}$/pxl $\times$ 0.0125$^{\prime\prime}$/pxl for a total field of view of 0.8$^{\prime\prime} \times$ 0.8$^{\prime\prime}$.   The data reduction is described in \cite{Mesa2019}. 
The phase of the central binary of the SINFONI observation is 0.459 \citep{Sissa2019}, and according to the interpretation of the light curve in that paper, the spectrum of the star should be dominated by the secondary at this epoch.

\begin{figure}
\centering
\includegraphics[angle=0,width=\hsize]{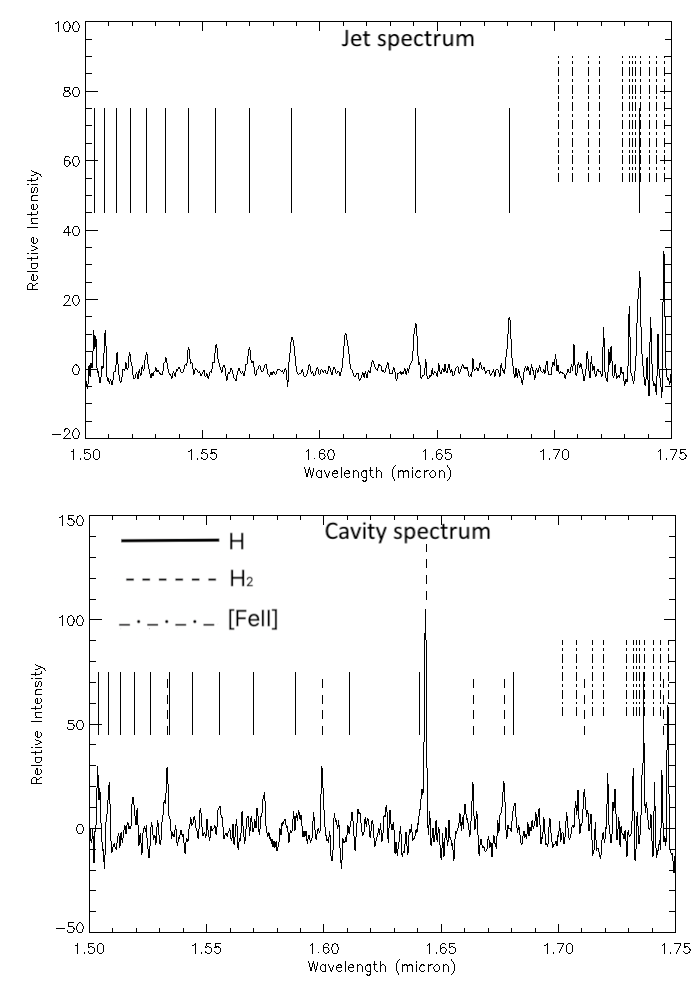}
\caption{Spectra from SINFONI data along the elongated structure for both "jet" and the "cavity" components. The locations of the Hydrogen recombination lines from the Brackett series are indicated, together with the [Fe~{\sc{ii}}] and H$_2$ lines. The spectra are extracted from the two boxes shown in Figure~\ref{Fig:composite_2}. } 
\label{Fig:spectra_sinfoni}
\end{figure}

In Figure~\ref{Fig:composite_2}, we show the composite line emission images obtained for the H, H$_2$ and [Fe~{\sc{ii}}] lines detected in the SINFONI data. Green refers to H-emission, blue to H$_2$-emission, and red to [Fe~{\sc{ii}}]-emission. While H and H$_2$ are almost co-located, the [Fe~{\sc{ii}}] emission shows a clear offset, being present at the southern edge of the "elongated structure" seen in the continuum image. In the same Figure, we identify two separated regions: one indicated as "jet" along the "elongated structure" direction, and the other one indicated as "cavity", which extends along the east direction. These regions were chosen arbitrarily, in a region where the bulk of the line emission is observed.
We extracted the spectra in these two regions, and they are shown in Figure~\ref{Fig:spectra_sinfoni}.  The spectra were obtained after dividing the total intensity of each wavelength for a synthetic spectrum of telluric absorption lines, and after  subtracting the median continuum emission. The synthetic spectrum of telluric absorption  was computed using the synthetic sky modeler, Telfit (\citealt{Gullikson2014}), adopting the proper parameters for airmass, pressure, humidity, and temperature. The Brackett series Hydrogen recombination lines, as well as H$_2$ emission lines, and [Fe~{\sc{ii}}] lines, are identified in the spectra.
A faint signal may be detected in Figure~\ref{Fig:spectra_sinfoni} in the direction of the counter-"elongated structure". This is most likely not a totally "elongated structure"-related signal, but it is contaminated by the typical butterfly pattern in the direction of the wind sometimes observed in in high-contrast images \citep{Milli2017}. To assess this hypothesis, we checked the wind direction for the night when the SINFONI observations were taken on the Paranal Ambient Conditions Archive, that is indeed $\sim$140$^{\circ}$ (counted clockwise from north), in agreement with the direction of the signal from "elongated structure" and counter-"elongated structure".

\subsection{FEROS data}

\begin{figure}[h]
\centering
\includegraphics[angle=0,width=9cm]{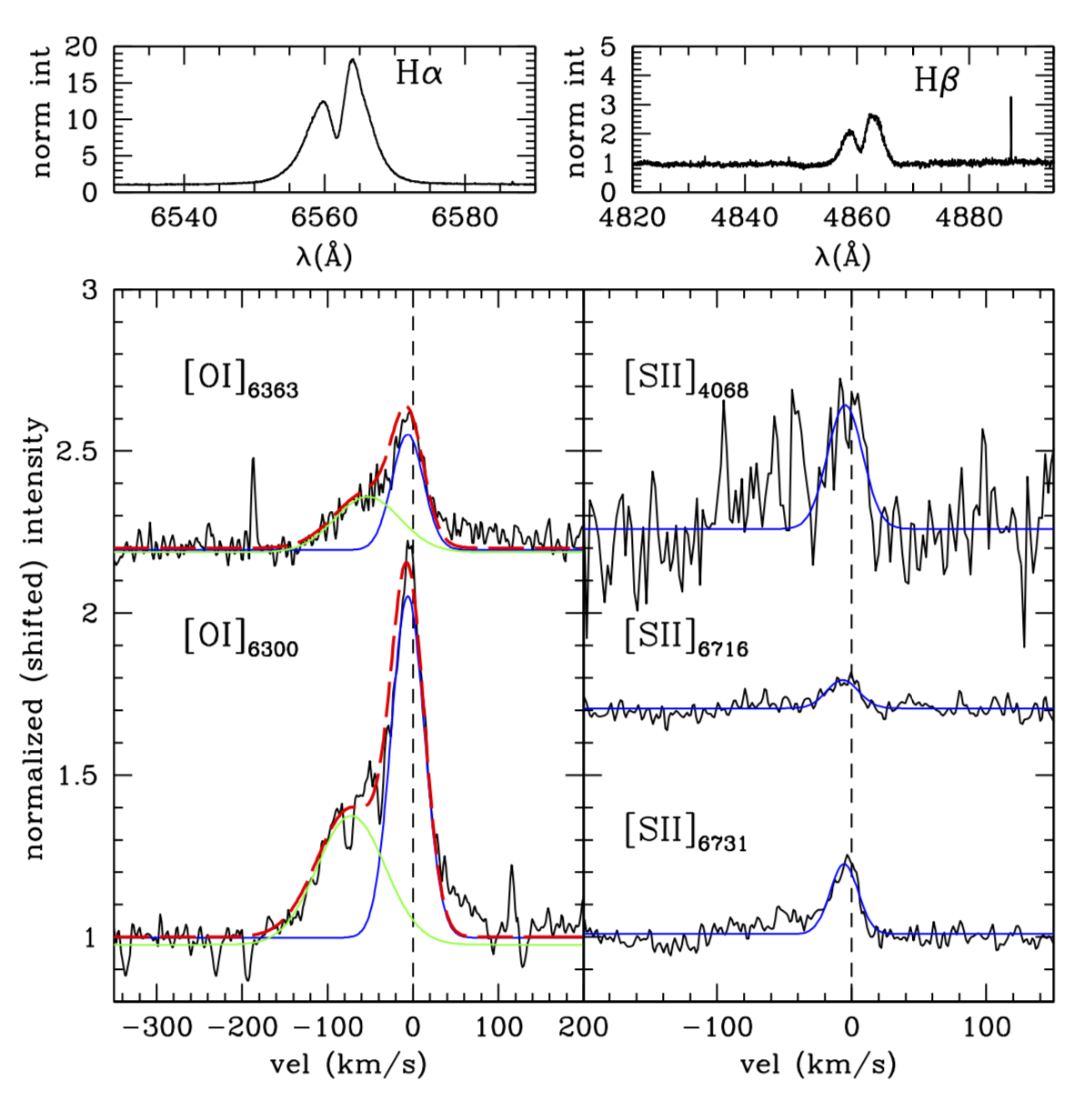}
\caption{FEROS spectrum of RCrA. Top panels: regions around H$\alpha$ and H$\beta$ lines. Bottom panels: regions around [O{\sc{i}}]$\lambda$6716/6731 and [S~{\sc{ii}}]$\lambda$4068/6716/6731 forbidden lines. The colored Gaussian profiles in the bottom panels refer to the multiple component deconvolution: in green, the HVC, in blue, the LVC, in red, the sum of the two components. } 
\label{Fig:forbidden_lines}
\end{figure}

We retrieved an optical spectrum of R~CrA from the ESO data-product Archive. It was acquired in 2009 with FEROS as part of the program 083.A-9013(A) and covers the wavelength range from 3500~\AA\ to 9200~\AA\ with a resolution of R$\simeq$48000 ($\simeq$6~km s$^{-1}$).  
The FEROS fiber diameter projected on the sky is 2.7\arcsec, and covers radii out to $\sim$200~au at the distance of R~CrA. The spectrum was reduced by performing flat-fielding, wavelength calibration, and barycentric correction. 
Assuming the $\sim$66-day period of the central binary system \citep{Sissa2019}, at the epoch of the FEROS observation, the spectrum was dominated by the primary star. 
We identified several emission lines in the spectrum of R~CrA, and we discuss the emission spectrum in the next section. 
 
\section{Data Analysis}

The analysis of the images and spectra described in the previous section allows us to investigate the extent, morphology, and physical condition of the gas and dust in the region where the "elongated structure" is observed.  
The detected lines and continuum probe different components of this complex system and allow us to derive the properties of the jet, the cavity, and the accreting gas, as explained in the following sections.

{\subsection{Gas properties}}

\begin{table*}
\caption{Observed lines in the FEROS spectrum}
\label{table:OIparam} 
\centering            
\begin{tabular}{lc|ccc|ccc|}
\hline\hline
Element  & Wavelength & \multicolumn{3}{c}{LVC} & \multicolumn{3}{c}{HVC}\\
         &            & v$_c$ & FWHM & EW & v$_c$ & FWHM & EW \\
         & (\AA)          & (km~s$^{-1}$) & (km~s$^{-1}$) & (\AA) & (km~s$^{-1}$) & (km~s$^{-1}$) & (\AA) \\
\hline      
$[$O~\sc{i}$]$ & 6300  & -5.8$\pm$0.2 & 45.6$\pm$0.7 & 1.22$\pm$0.01 & -71.8$\pm$3.7 & 94.9$\pm$6.8 & 0.40$\pm$0.01\\
$[$O~\sc{i}$]$ & 6363 &  -6.2$\pm$0.4 &43.2$\pm$1.4 & 0.50$\pm$0.01  & -50.4$\pm$7.8 & 94.7$\pm$14.9 &0.16$\pm$0.009\\
$[$S~\sc{ii}$]$ & 6731 & -5.6$\pm$0.6 & 24.3$\pm$1.7 & 0.18$\pm$0.01 & ... & ... & ...\\
$[$S~\sc{ii}$]$ & 6716 & -7.6$\pm$6.1 & 28.5$\pm$2.9&  0.093$\pm$0.009 & ... & ... & ...\\
$[$S~\sc{ii}$]$ & 4068 & -6.8$\pm$4.1 & 36.0$\pm$12.6 & 0.28$\pm$0.02 & ... & ... & ...\\
$[$Fe~\sc{ii}$]$ & 7155 & ... & ... & ... & -35.8$\pm$1.4 & 129.8$\pm$4.5 & 0.26$\pm$0.01 \\
\hline      
\end{tabular}
\end{table*}

The FEROS spectrum covers a wavelength range rich in emission lines that are diagnostic of accretion and outflow activity. Strong, double-peaked, self-absorbed H$\alpha$ and H$\beta$ emission lines are observed, together with atomic forbidden lines such as [O~{\sc{i}}] and [S~{\sc{ii}}]. 

The ratio between the intensity of the lines of  [S~{\sc{ii}}] at 6716 and 6731~\AA\ has been largely used to derive the electron density $n_e$ of the emitting gas, because they have a similar energy of the upper level. The intensity ratio is then sensitive to electron density and almost independent of the temperature \citep{Osterbrock1989}. We used the revised diagnostic diagrams from \cite{Proxauf2014}, which give  $n_e$=7$\times$10$^{3}$cm$^{-3}$ for the [S~{\sc{ii}}]~$\lambda$6716/6731 line ratio, assuming an electron temperature of 10,000~K. This value is in agreement with the value found from the intensity ratio of the [S~{\sc{ii}}]~ lines at 4069 and 6731~\AA\  by \cite{Hamann1994}. 

The high resolution of the FEROS spectrum also allowed us to conduct an analysis on the multiple component of the Oxygen forbidden lines. Emission from [O~{\sc{i}}] lines in the optical wavelengths is a well established tracer of outflows in T Tauri stars. Several studies of these lines have shown that their emission is often blueshifted and is formed in an outflow whose redshifted part is obscured by the circumstellar disk (e.g. \citealt{Edwards1987}). Moreover, the [O~{\sc{i}}] lines observed at medium/high-resolution show a profile composed of multiple components: a high-velocity component (HVC) with line peaks shifted by up to a few hundreds~km~s$^{-1,}$  and a low-velocity component (LVC) with blueshifts between a few to $\sim$30~km~s$^{-1}$. The line components are emitted in physically different regions. The HVC is produced in a fast-moving collimated (micro) jet (e.g. \citealt{Kwan1988, Hartigan1995}), while the LVC has been found to most likely trace disk winds (e.g. \citealt{Acke2005, Rigliaco2013, Natta2014, Banzatti2019}). In Fig.~\ref{Fig:forbidden_lines}, we show the [O~{\sc{i}}] 6300~\AA\ and 6363~\AA\ line profiles of R~CrA, and in Table~\ref{table:OIparam}, we summarize the forbidden line properties. Both lines clearly show the two components that can be reproduced by Gaussian profiles: in blue the LVC, blueshifted by $\sim$~6km s$^{-1}$, and in green the HVC tracing gas moving at higher velocity. The observed profile is well reproduced by the sum of two Gaussian components (red profile).  These features are observed within the 2.7$^{\prime\prime}$ FEROS fiber, meaning that the gas is distributed within 200~au from the central star. This represents the only solid limit we can put on the region where the HVC form as seen on-sky. More speculatively, if we assume that these lines are emitted from gas in Keplerian orbit around the star, we can use the measured FWHM (see Table~\ref{table:OIparam}) to constrain the radius at which they are emitted. Using the sum of the masses of the two components of the central binary system, respectively, 3.0~M$_{\odot}$ and a 2.3~M$_{\odot}$, we find that under this hypothesis, the [O~{\sc{i}}] LVC should be emitted at $\sim$2.5~au from the central star. On the other hand, the LVC can be either magnetically or thermally driven, and in both cases, its origin is within few au from the central star. Following \citealt{Hartigan1995}), we used the [O~{\sc{i}}] 6300~\AA\ HVC line luminosity to retrieve an estimate of the mass loss through the jet, to be compared then to the mass accretion rate onto the star. Using equation (A8) from \citealt{Hartigan1995}), assuming $l_{\bot}\sim$1.0~$^{\prime\prime}$ as the jet aperture on the plane of the sky (as retrieved from the images), and $v_{\bot}$ the projected velocity of the jet in the sky ($\sim$760~km/s, as discussed in the next section), we retrieved $\dot{M}_{loss}\sim$2.8$\times10^{-7}$M$_\odot/yr$. This value is then compared to the accretion rate we retrieved from the H$\alpha$ and H$\beta$ lines. We measured the equivalent width (EW) for these lines, and obtained an estimate of the accretion luminosity L$_{acc}$ of the star using the empirical relationships given by \citet{Fairlamb2017} for Herbig Ae/Be stars. To overcome the uncertainty caused by variability of the star, we employed the photometry taken the same night as the FEROS spectrum (V$_{mag}$=11.6) with the All Sky Automated Survey (ASAS, \cite{Pojmanski1997}) for these measurements. 
The measured EW$_{H\alpha}\sim$135\AA, and  EW$_{H\beta}\sim$11\AA, yield to accretion rate values of 1.6$\times10^{-6}$M$_{\odot}$/yr and 5.1$\times10^{-6}$M$_{\odot}$/yr, respectively. Assuming the average accretion rate is the mean between these two values, we find $\dot{M}_{acc}\sim$3.3$\times10^{-6}$M$_{\odot}$/yr. The ratio of the mass accretion to the mass loss rate is $\sim$0.08, which is consistent with values found for TTSs with prominent stellar jets (e.g., HH34, HH47, HH111 \citealt{Hartigan1994, Hartigan1995}), and it is consistent with a jet that is powered by the accretion onto the central star. 

As we mentioned in the previous section, the SINFONI spectrum in the "jet" region covering the near-IR wavelengths range from 1.45~$\mu$m to 1.85~$\mu$m, does show the Brackett series Hydrogen recombination lines, but there is no detection of [Fe~{\sc{ii}}] lines at 1.53, 1.60 and 1.64~$\mu$m in this region. Studies of jets from Herbig stars (as also recently found for RY Tau, \citealt{Garufi2019}) in the optical and in the Near-IR showed that optical jets also emit in the Near-IR  [Fe~{\sc{ii}}] lines, unless the density in the jet beam is very low ($n_e$<10$^{3}$cm$^{-3}$). Indeed, owing to their large critical density, these lines need high electron density to be efficiently excited, thus they cannot be used as diagnostics in low-density jets (e.g., \citealt{Podio2006}). Besides being caused by the low-density gas, the non-detection of the [Fe~{\sc{ii}}] might be due either to the low sensitivity of the H-band spectrum, or to the depletion of the iron that is locked into the dust grains (e.g., \citealt{Podio2006, GarciaLopez2010}).

\subsection{"Elongated structure" size} 

\begin{figure}[h]
\centering
\includegraphics[angle=0,width=8cm]{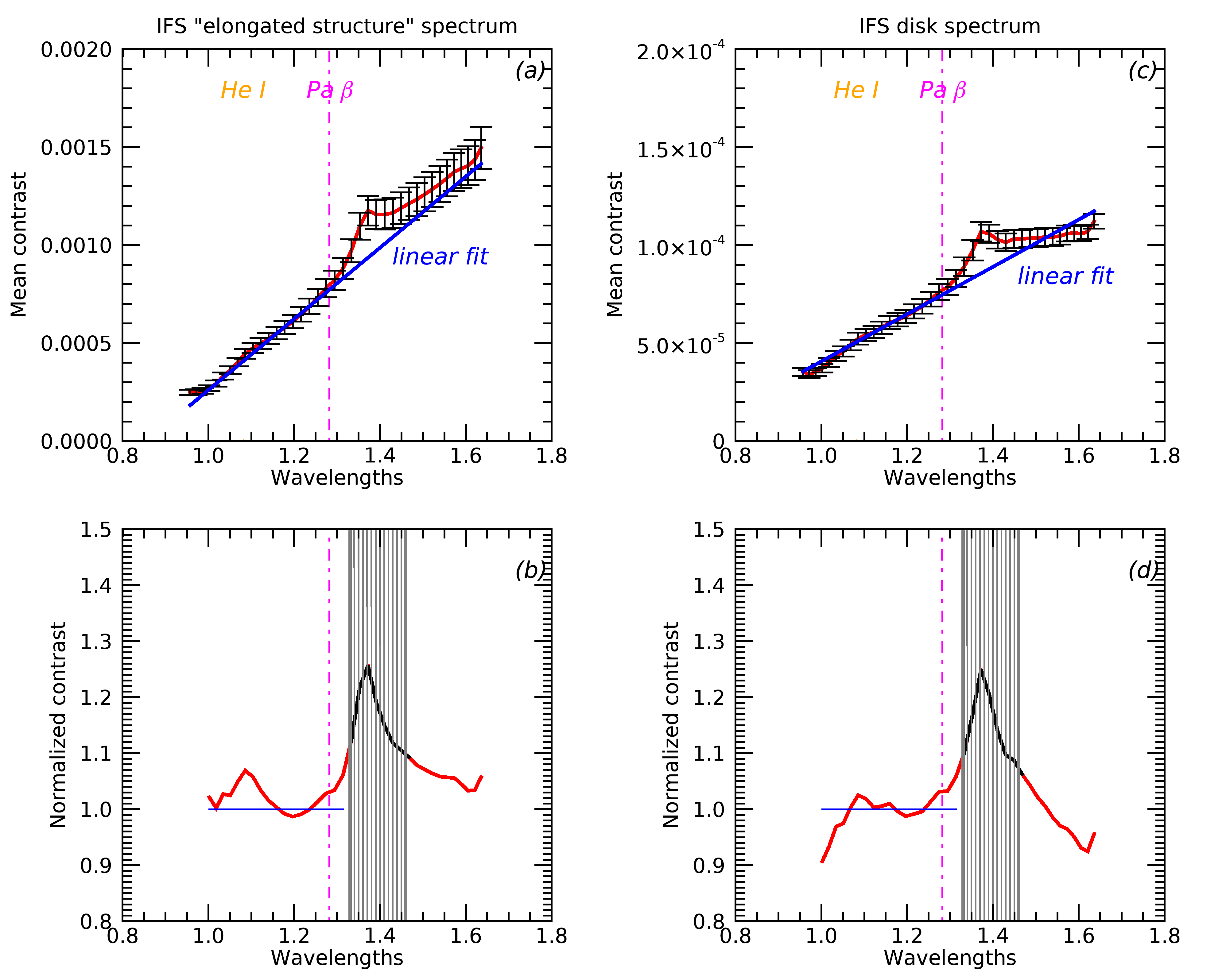}
\caption{Spectra extracted from SPHERE-IFS data, once a radial profile, has been subtracted. Top left panel (a):  spectrum (mean contrast versus wavelengths) extracted in area where"elongated structure" is identified, normalized for stellar peak. The location of the Helium and Hydrogen (Pa$\beta$) line emission are labeled. The blue line shows the linear curve utilized to normalize the spectrum, and reported in the bottom panel. Bottom left panel (b): spectrum normalized to linear fit indicated in top panel. The grey-shaded area shows the telluric band. Top right panel (c): same as panel (a), but for a region closer to the disk identified by \citet{Mesa2019}. Bottom right panel (d): same as panel (b) for disk.} 
\label{Fig:IFS_spectrum}
\end{figure}

The SPHERE images of R~CrA (IRDIS and IFS) obtained using the monochromatic PCA with 1 principal component are shown in Figures~\ref{IFS_total} and ~\ref{IRDIS_total}. 
In Figure~\ref{Fig:IFS_spectrum}, we show the median spectrum from the IFS data cube. It represents the average signal within the area where we identify the "elongated structure"  normalized to the stellar peak. The spectrum is dominated by continuum emission: however, there is some evidence of the contribution due to the emission line of Helium at 1.083~$\mu$m, and slight hints of Paschen $\beta$\ when the spectrum is normalized to the continuum (Fig.~\ref{Fig:IFS_spectrum}, left bottom panel). The steepness of the observed spectrum shows that the "elongated structure" continuum emission has a redder color than the stellar emission. 
For comparison, we also extracted the spectrum in the region where the almost edge-on disk was pointed out by \cite{Mesa2019} (panels (c, d) in Fig.~\ref{Fig:IFS_spectrum}). We notice that while the reddening of the spectrum in the direction of the disk is in agreement with the spectra extracted for other disks around Herbig stars (e.g., HD~100546, \citealt{Sissa2018} and SAO~206462, \citealt{Maire2017}), the spectrum along the "elongated structure" appears steeper.  
We must consider that the circumstellar environment around RCrA is affected by high extinction (e.g., \citealt{Bibo1992, Sissa2019}), and a gradient of extinction within the region itself, hence the reddening of the spectrum might be due to this effect. However, the scattering properties of dusty grains of different size might also play a role in this context. 

We analyze the aperture of the "elongated structure". Following on from the analysis of \cite{Mesa2019}, we measured the aperture from the IFS continuum image (Fig.~\ref{IFS_total}, top panel). The position angle (PA) of the "elongated structure" spans from $\sim$30$^{\circ}$ to $\sim$70$^{\circ}$, with a median PA of $\sim$50$^{\circ}$. The same PA and "elongated structure" aperture is shown in the IRDIS image, where the contour plot of the "elongated structure" is also shown. The furthest region of the "elongated structure" marked by the contours appear slightly bent toward south.  
We notice, moreover, that the dust "cavity" region, as identified in the SINFONI image, extends outside the "elongated structure" as identified from the IFS and IRDIS continuum and line images. In particular, the axis of the emission in the dust "cavity" region has a PA of $\sim$100$^{\circ}$.

The continuum emission of the "elongated structure" observed in the IFS image, which represents the approaching part to the observer, covers all the IFS field of view, meaning that it extends up to $\sim$120~au from the central system. From the IRDIS image, the radial extent of the approaching part of the observed "elongated structure" extends up to 2.6$^{\prime\prime}$ from the central objects (400~au at the R~CrA distance), remaining as wide as $\sim$30$^{\circ}$\ up to $\sim$300~au. Optically visible jets from TTSs are known to begin with wide (10-30~degree) opening angles close to the source, and are rapidly collimated to within a few degrees in the innermost 50-100~au \citep{Ray2007, Frank2015}. Wider structures, as the one observed in the continuum emission around R~CrA, might be consistent with shells of ambient gas swept up by the jet bow-shock and a surrounding slower wider-angle component. This wide-angle wind and the swept-up outflow expand more slowly, carving out a cavity, which widens over time into the envelope and the surrounding cloud \citep{Frank2015}.

\subsection{Morphology}

\begin{figure}
\centering
\includegraphics[angle=0,width=\hsize]{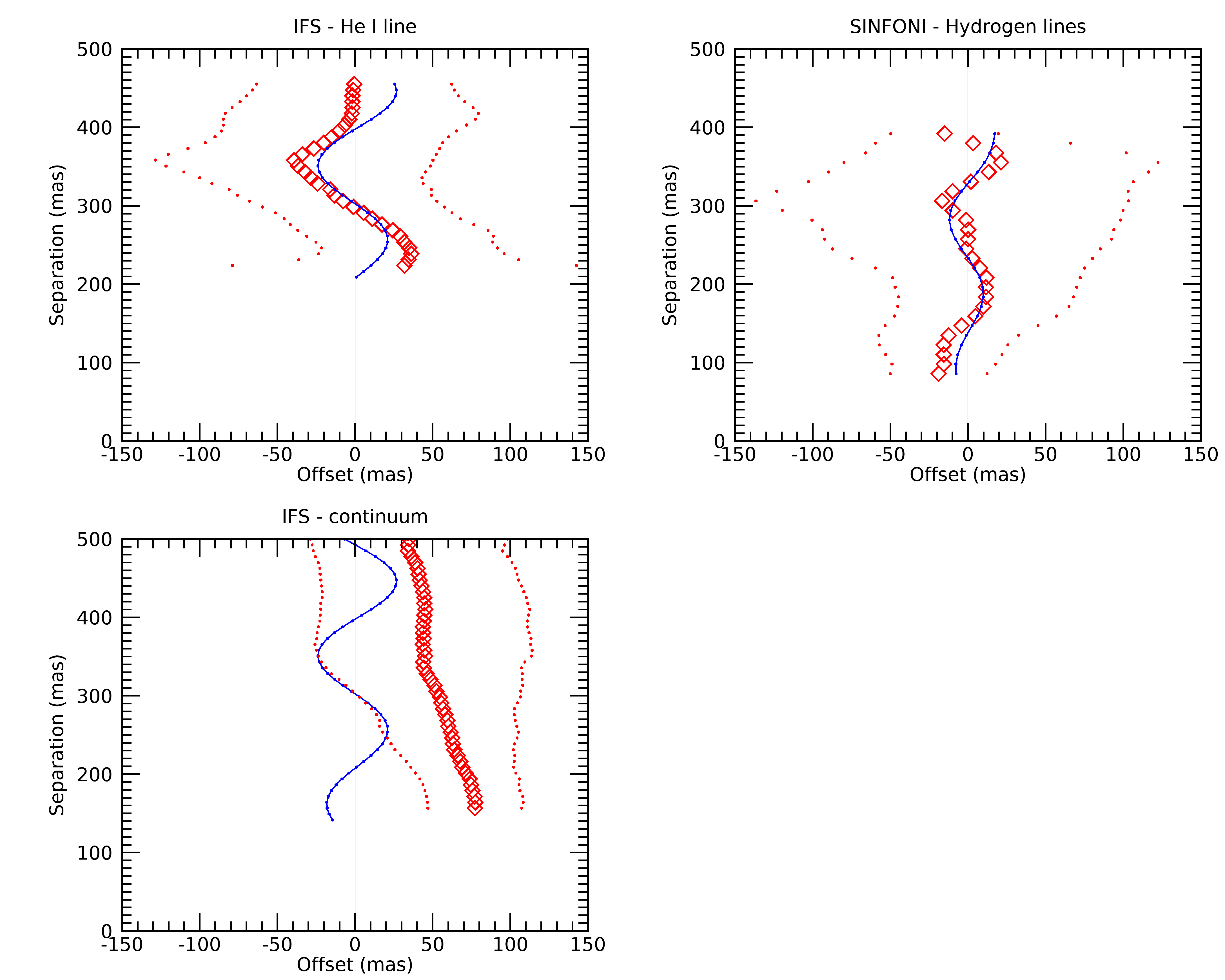}
\caption{Peak position of Hydrogen and Helium lines from SINFONI and SPHERE-IFS data, respectively, as a function of distance from central objects. Dots indicate the sigma of the Gaussian distribution used to obtain the peak position within the slice. The solid blue line is the fit of the wiggle if it is produced by the orbital motion of the jet around the binary system at the center. The blue fit in the three panels is the same, however in the panel showing SINFONI data it is adjusted in order to account for the difference in phase between SPHERE and SINFONI observations and for the different instrument resolution. } 
\label{Fig:wiggle_lines}
\end{figure}

The overall morphology of the observed "elongated structure" as seen in the continuum emission from IFS and IRDIS images  appears non-uniform and discontinuous along the radial extent. Moreover, the SINFONI image shows a wiggling pattern in the Hydrogen lines. 
We analyzed the wiggling as a function of the distance from the central binary by focusing both on the emission lines and on the continuum emission. We employed the same method used by \cite{Antoniucci2016}: we considered a set of contiguous slices orthogonal to the jet axis and in each slice we fitted the pixel distribution with a Gaussian function in order to obtain the profile peak positions as a function of the distance from the star. We applied this method to the SPHERE-IFS channels containing the He~{\sc{i}} line at 1.083$\mu$m, to the SINFONI channels containing the Brackett series lines at 1.555$\mu$m, 1.641$\mu$m, 1.681$\mu$m, 1.736$\mu$m, and to the IFS average continuum emission. The results are shown in Fig.~\ref{Fig:wiggle_lines}, where the plots from the gas components are shown in the top panels, and the continuum emission is shown in the bottom panel. We first notice that the "elongated structure" seen in the gaseous component (both Helium and Hydrogen) is at a median PA of $\sim$65$^{\circ}$, higher than the median PA ($\sim$50$^{\circ}$) found from the continuum emission, pointing to a misalignment between the "elongated structure" axis seen in the continuum and the gas emission. 
This misalignment can already be clearly noted in Fig.~\ref{IFS_total}, where the He~{\sc{i}} emission (bottom panel) does not share the same PA as the continuum emission (top panel). The same happens for the Hydrogen, [Fe~{\sc{ii}}]  and H$_2$ lines shown in Figure~\ref{Fig:composite_2}, where the lines are not emitted at the same PA as the continuum in Y-H and K bands from the IFS and IRDIS images.
Both Helium and Hydrogen emissions display a wiggling pattern with a projected half-opening angle of 5$^{\circ}$ -- 6$^{\circ}$. It is not surprising to see a wiggle in the jet of young stars: it has been detected in both the other young stars where a jet was observed with SPHERE, namely RY Tau \citep{Garufi2019} and ZCMa \citep{Antoniucci2016}. Contrary to these previous studies, where the binarity of the central object was assumed from the wiggle of the observed jet, for R~CrA we know that the central system is a binary star \citep{Sissa2019}. We checked that the rotation period of the binary system (66 days) is in agreement with the phase difference between the two observations, SINFONI and SPHERE-IFS, which are shifted by $\sim$two months in time. If the observed period of the wiggling (190~mas) is linked to the period of the central binary system, we could then measure the velocity of the jet projected on the sky, which we found to be $\sim$760~km s$^{-1}$. This corresponds to a radial component of the velocity respect to us of $\sim$130~km s$^{-1}$, if we consider that the disk is seen almost edge-on and that the jet is then seen with an inclination of $\sim$10$^{\circ}$ on the plane of the sky (see below). 
We also notice that the wiggling is not visible in the continuum emission. As shown in the bottom panel of Fig.~\ref{Fig:wiggle_lines}, the model obtained from the emission lines (blue solid line), using the same PA measured from the gas lines, does not reproduce the shape of the observation, pointing to the conclusion that the continuum emission does not have a wiggling pattern, even if is appears non-uniform.  

Another point coming from the analysis of the SPHERE-IFS images in the different channels is that the emission by the gas shown by He{\sc i} line is not centered on the continuum emission. This is clear comparing the top and bottom panels of Fig.~\ref{IFS_total}: the He{\sc i} emission is not exactly co-located to the continuum emission, appearing shifted to the south direction, almost at the border of the continuum emission, suggesting it is produced in the external layer of the "elongated structure". Moreover, the He{\sc i} emission is spatially resolved along the direction perpendicular to the jet elongation, yielding a median jet semi-aperture of $\sim$10~au. 

\subsection{Geometrical model of the wiggling jet}

\begin{figure*}
\centering
\includegraphics[angle=0,width=\hsize]{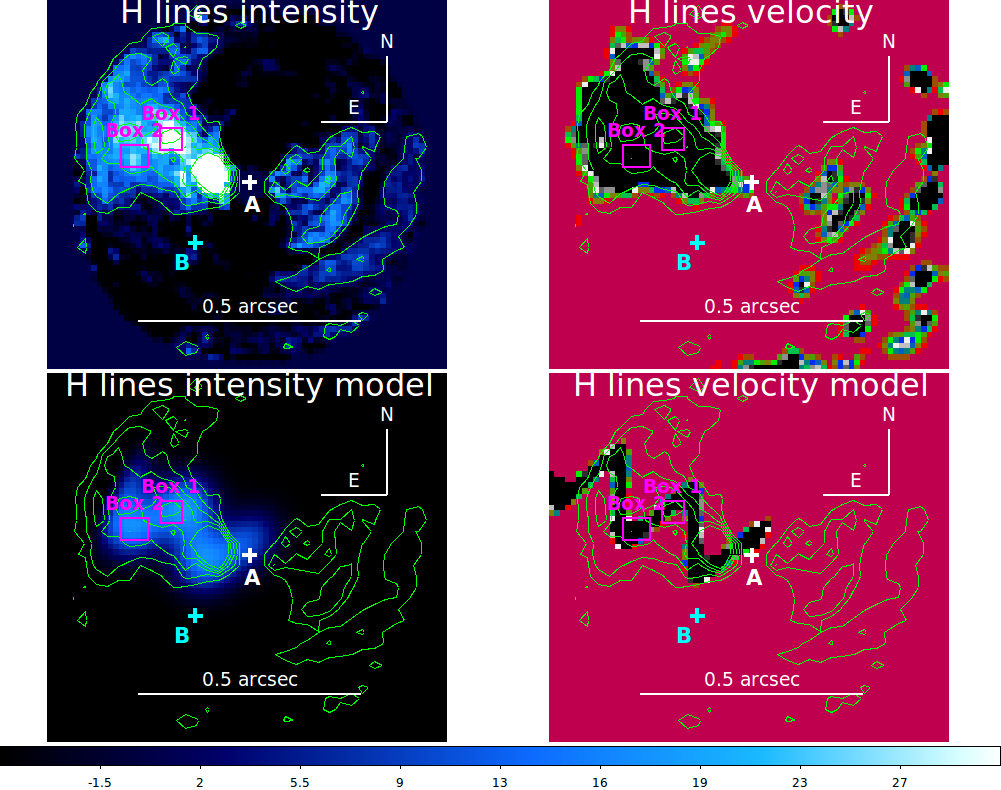}
\caption{Top panels: Intensity and velocity map of HI lines obtained from SINFONI data. The intensity map is obtained by measuring the sum of the intensity of the HI lines in each pixel, and the velocity map is obtained from the cross-correlation function of the image using a correlation mask. Bottom panels: intensity and velocity maps of model. Box1 and box2 identify the two consecutive blobs that are used to build the geometrical model of the wiggling, as discussed in the text.}
\label{velocity_model}%
\end{figure*}

We constructed a geometrical model of the wiggling jet of R~CrA, as observed in the H{\sc i} lines in the SINFONI spectrum. We modeled two observed quantities: the luminosity and the radial velocity maps. The radial velocities were derived by cross-correlation, and the maps of the intensities and radial velocities of the H{\sc i} lines are shown in the top panels of Figure~\ref{velocity_model}. 
In order to model these two quantities, we assume that the jet is describing a helix on the surface of a truncated cone, which is seen with an inclination $i$\ and position angle $PA$\ with respect the sky plane. Since the inclination is very small, the jet velocity on the sky plane is set by the ratio of the separation between the helix pitch and the binary orbital period. In the  assumptions that the jet is optically thin, the regions where the spiral crosses the plane of the sky have a greater depth along the line of sight, hence they are seen as blobs in the luminosity image. We identified two consecutive blobs in Figure~\ref{velocity_model} as box1 and box2.  The projected separation between two consecutive blobs on the same side of the jet will then give the helix pitch. As we have seen previously, the jet transverse velocity determined in this way is 760 km s$^{-1}$.
The combination of the luminosity and radial velocity maps make it possible to determine all the parameters of the model (see Figure~\ref{velocity_model}). The PA is the mid-angle of the jet on the sky plane (PA=67${^{\circ}}$, in the usual convention starting from north toward east). The ratio between the mean jet radial velocity and the transverse velocity sets the inclination angle at which the helix is seen. We find $i$=3.3${^{\circ}}$, meaning that the jet is seen very close to the sky plane. This suggests that the disk of R~Cra is seen almost edge on. In this model, the radial velocity difference between opposite sides of the helix (e.g. box 1 and box 2 in Figure~\ref{velocity_model}) is due to a projection effect, because they are moving with a different inclination with respect to the line of sight, and then  sets the cone aperture angle. The observed difference of $\sim$18~km s$^{-1}$ is reproduced by a semi-aperture of the cone of $\sim$1.2${^{\circ}}$, pointing to a very well collimated jet. Once this angle is fixed, the observed distance between the blobs and the jet axis determines the radius of the base of the cone $r_B\sim$5.3~au.

Most interestingly, the model also sets the phase when the jet crossed the binary orbital plane at the epoch of the SINFONI observation (at a distance of 5.3~au from the barycenter of the central binary). Of course this is an idealization of the real jet trajectory because it is unlikely that the jet is launched from this position. However, this can still be useful in  understanding the jet geometry with respect to the central binary. We found that this phase is 0.35, where the zero point is when the binary primary (R~CrA Aa) is in opposition. For comparison, the binary model constructed in \citet{Sissa2019} tells us that at the same epoch, the R~CrA Aa was at a phase of 0.459. The difference between these two phases is quite small, and may be justified by considering the delay of the jet due to the time required to reach a distance of 5.3 au from the star at a constant speed of 700 km s$^{-1}$, 12 days, that corresponds to a phase difference of 0.18. Of course, the jet trajectory is likely shorter than one passing through this ideal point. 
Even if these values contain some uncertainties (e.g. in the exact jet trajectory and velocity and in the exact value of $r_B$), they suggest that the jet is actually launched along the direction defined by the two components of the central binary, towards the far side of the primary. 

\begin{table}
\caption{Radial velocities and FWHM for spectral lines detected in SINFONI spectra. $v_c$ and FWHM were derived by cross-correlation and fitting of the following lines. 
Hydrogen: Brackett series lines from H(11-4) up to H(21-4); 
$[$Fe{\sc ii}$]$: 1.53347, 1.59948, 1.66377, 1.67688, 1.711132; H$_2$: 1.70150, 1.71427, 1.72902, 1.73202.  }
\centering
\begin{tabular}{lcc}
\hline
\hline
Component   &     $v_c$     &     FWHM      \\
            & (km s$^{-1}$) & (km s$^{-1}$) \\
\hline
\multicolumn{3}{c}{Star}\\
\hline
H           &     -14.8$\pm$4.0 &      435$\pm$14\\
\hline
\multicolumn{3}{c}{"Jet"}\\
\hline
H            &   -39.9$\pm$2.4  &    393$\pm$ 7\\
H (box 1)    &   -29.7$\pm$3.0  &     383$\pm$ 9\\
H (box 2)    &   -45.3$\pm$2.7  &     410$\pm$ 9\\
H$_2$           &   +0.7$\pm$4.9  &     174$\pm$12\\
\hline
\multicolumn{3}{c}{"Dust Cavity"}\\
\hline
$[$Fe{\sc ii}$]$ & -0.2$\pm$4.5  &     208$\pm$12\\
H            &    -6.6$\pm$8.1   &       423$\pm$26\\
\hline
\end{tabular}
\label{tab:sinfoni}
\end{table}

\subsection{[Fe~{\sc ii}] and H$_2$ lines}

Table~\ref{tab:sinfoni} collects radial velocities and full width at half maximum (FWHM) of the spectral lines detected in the SINFONI spectra. In particular, we collected data for spectra extracted from different regions of the data cube: close to the star, and in the "jet" and dust "cavity" area, as identified in Fig.~\ref{Fig:composite_2} and in Sect.~2.2. The Hydrogen lines are detected in all regions and are usually very broad, consistent, and with a high velocity wind. The [Fe~{\sc ii}] are only detected in the dust "cavity" area, tracing an external layer of the "elongated structure". They have low velocities, and are narrow and not resolved in the SINFONI spectrum. The ratio between the intensity of the lines at 1.64 and 1.53~$\mu$m is a diagnostic of density \citep{Nisini2005}. The observed ratio of about four corresponds to a density of about 10$^4$~cm$^{-3}$. This is an order of magnitude larger than the upper limit obtained from the lack of [Fe~{\sc ii}] lines in the "jet" region, but not far from the one estimated using the [S~{\sc ii}] lines and the narrow component of the [O~{\sc i}] lines. This suggests that this dust "cavity" region corresponds to the low-velocity component seen in the FEROS spectra (see Table~\ref{table:OIparam}). 

H$_2$ lines are also clearly detected in the SINFONI spectra, and their emission comes from a region very close to the star, as seen in Fig.~\ref{Fig:composite_2}. Even if an exact quantification is difficult, it suggests that it originated within 230 mas (corresponding to $\sim$35 au from the star). This is within the orbit of the M-dwarf companion, where the circumbinary disk should be. 
The radial velocities of these lines are in agreement with those derived in the Hydrogen lines in the "jet" direction, but the FWHM are much smaller, and the lines are not resolved in the SINFONI spectrum. 

The relative intensity of the different H$_2$ lines agree, within 10\%, with what was expected from collisional excitation at a temperature of about 15000~K, and it is very different from what was expected from fluorescence models \citep{Black1987}. Namely, mainly low-excitation lines are observed, and the high-excitation lines, which are also expected to be strong due to fluorescence, are absent. The high-temperature gas can be heated by shocks \citep{Shull1982} that may be located close to the base of the jet or in the regions where this interacts with the disk. 
Radial velocities and FWHM of the Hydrogen and H$_2$ lines in the "jet" area are in agreement with the values found for the [Fe~{\sc ii}] at 7155~\AA\ line in the FEROS spectrum, suggesting that this line (that requires densities as high as  $\sim$10$^{5}$cm$^{-3}$, \citealt{Nisini2005}) might be also produced in the same post-shock region near the star. 

We notice that the H$_2$ lines, which should dominate the K1-band in the IRDIS image, are not detected.  To stress this point, we performed the subtraction of the K1-K2 IRDIS image, as shown in  Fig.~\ref{IRDIS_total} (bottom panel).  On the other hand, fainter H$_2$ lines are detected in the H-band SINFONI spectrum. This non-detection of H$_2$ in the K1-K2 image is essentially due to the fact that SPHERE has been designed for detecting continuum emission, it is not optimized for detecting (extended) line emission. The spectral resolution of SINFONI is two orders of magnitude higher: this makes it possible to detect emission lines with a surface luminosity two orders of magnitude fainter. We verified that the non-detection if the brightest H$_2$ emission line in the K1-band (at 2.12$\mu$m, (S(1)(1-0))) is compatible with the detection of the brightest H$_2$ emission line in the H-band (at 1.74$\mu$m, (S(7)(1-0))). 

\subsection{Spectro-astrometry: information on the verycentral regions}

Some information about the very central region of the system can be obtained using spectro-astrometry \citep{Whelan2008} on the SINFONI data. We should remind readers that the bulk of the emission from R~Cra in the near-IR is due to the warm disk \citep{Sissa2019}. Interferometric observations with the instrument AMBER at VLTI (\citealt{Kraus2009}) indicate that this emission is offset by about four mas with respect to the barycenter of the system, in the direction of the jet, because self-shadowing of the highly inclined disk only makes it possible to see the far side of the disk. Also, the separation between the central binary components projected along the jet axis is very small (about 0.3-0.4 mas) and can thus be neglected here. Since the scope of spectro-astrometry is to explore regions very close to the center of the system, we used the images obtained by subtraction of a radial profile. The images were corrected for the telluric lines and then cross-correlated with digital masks for HI and [Fe~{\sc ii}] (results for H2 lines were less clear). The resulting data cube is then made of cross correlation functions - that is, the z-coordinate now represents radial velocities. We then rotated these images so that the jet is along the y-axis, and collapsed the images along the jet. This is how we obtained bidimensional images, where the axes are radial velocity and offset along the jet with respect to the peak of the continuum emission. We then fitted Gaussians in offset as a function of radial velocity and compared the position of the peak with the value we obtained for the continuum (that, we remind readers, represents the position of the warm disk). Since there is no relevant variation of the position of this peak with velocity, we simply considered the average values. The peak obtained for HI lines is offset by about four mas in the opposite direction to the jet: this indicates that the bulk of the H emission is caused by material very close to the central binary, likely tracing the accretion. This agrees with the large value of the FWHM. On the other hand, the peak for [Fe~{\sc ii}] is offset by about two mas in the direction of the jet, indicating that, in this case, the bulk of the emission comes either from the jet or from regions close to the disk, at about six mas ($\sim 1$~au) from the star projected along the jet axis.

\section{Proposed scenario and discussion}

In this section, we discuss a scenario where all the features, structures, and properties seen and analyzed in the "elongated structure" are taken into account. The proposed scenario is supported by at least a similar case, the T Tauri star FS~Tau~B observed by \citet{Eisloffel1998}, who imaged the "elongated structure" as the wide outer edges of windblown cavities and the narrower jet flowing inside the cavity in H$\alpha$. 

\begin{figure}
\centering
\includegraphics[angle=0,width=9cm]{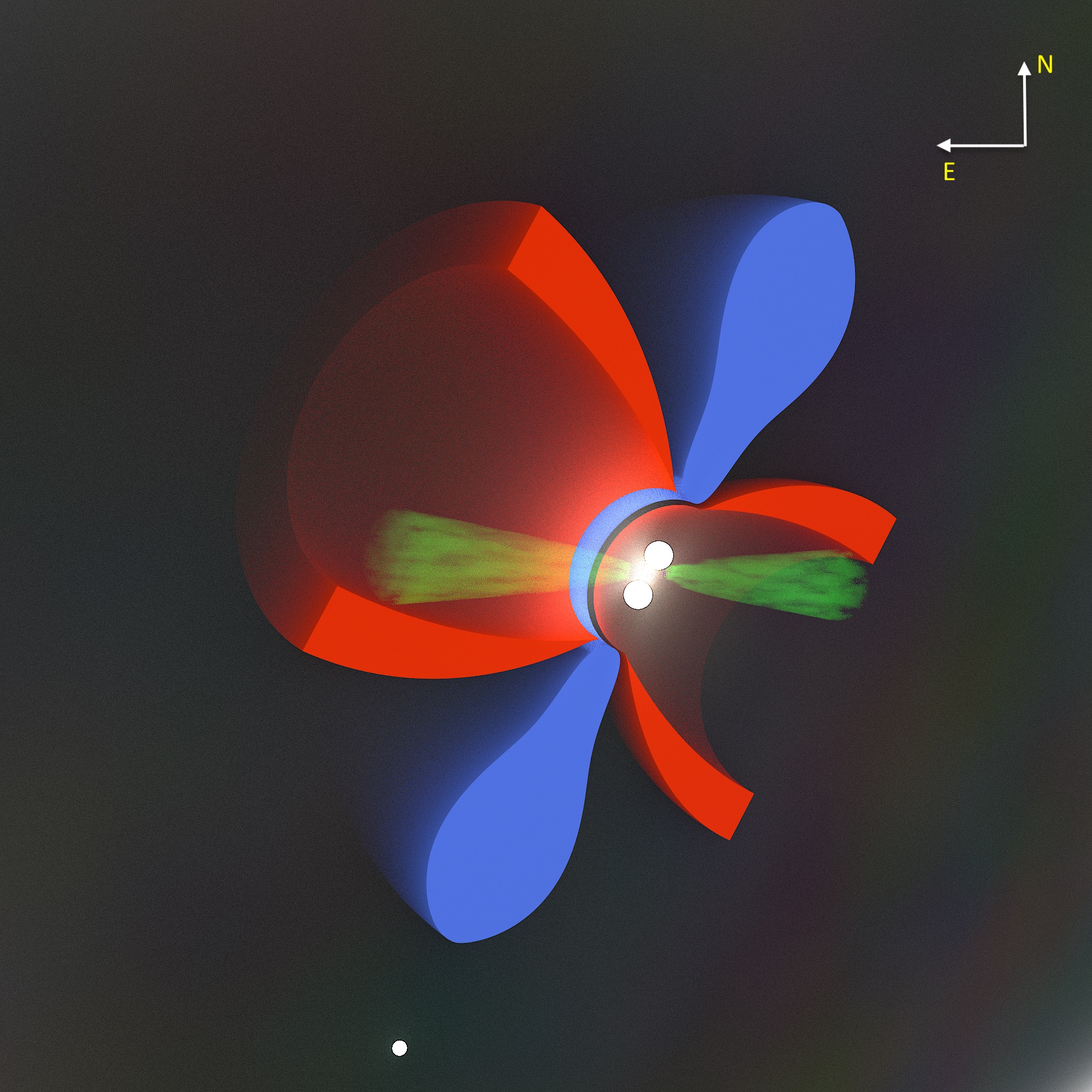}		
\caption{Sketch of scenario proposed for the "elongated structure" of R~CrA. White-filled circles are the three stellar components. The blue region is the circumbinary disk. The green area is the gaseous high velocity jet. The red region is the edge of the dust cavity. The stars of the triple system are represented with the white dots.} 
\label{Fig:sketch}
\end{figure}
 
The schematic picture we consider is obtained from the complementary detailed analysis of the SPHERE (IRDIS and IFS) images and spectra, of the SINFONI data, and of the optical FEROS spectrum. The system appears to be composed of at least two elements: a dusty component seen in scattered light, and a gaseous component, detected in emission atomic lines of forbidden species ([O~{\sc i}], [He~{\sc i}] and [Fe~{\sc ii}]) and Hydrogen lines. 
What we have called "elongated structure" until now, which is seen in scattered light in the IFS and IRDIS images, as well as in the SINFONI image, is dominated by continuum emission and is most likely a cavity carved out into the circumstellar environment; the dust on the cavity walls is illuminated by the central binary and scatters light toward us. 
The jet flowing inside the cavity is detected in the HVC of the optical forbidden lines and in the broad Hydrogen lines. The jet is very collimated and shows a wiggling pattern in the Hydrogen and [H~{\sc i}] lines, which is in agreement with the orbital period of the central binary. 
The LVC of the optical forbidden lines suggests the presence of gas moving more slowly, most likely a disk wind. SINFONI data allows us to spatially resolve the [Fe~{\sc ii}] emission, which appears to be located at the edge of the cavity, consistent with a layered disk wind as well. 
This structure is captured in the schematic picture of the R~CrA environment that is shown in Figure~\ref{Fig:sketch}. The disk, which is illustrated in blue in the Figure, is not well detected in the scattered light images, and it is most likely seen edge on. The central binary system \citep{Sissa2019} is hidden behind the coronagraph, while the wide M-dwarf companion is clearly visible in all the images.  The red shell shows the cavity walls, and the green diffuse area the gaseous jet flowing within the cavity.  

Differently from FS~Tau~B \citep{Eisloffel1998}, where the narrower jet flowing inside the cavity points toward the center of the cavity, in the case of R~CrA the gaseous jet is not currently pointing toward the center, but rather toward the southern border of the cavity. The misalignment between the axis of the cavity and the jet might be due to a temporal variation of the jet direction. In turn this might be attributed on short timescales to a wiggling pattern of the jet related to the central binary orbit, and on longer timescales to a precession motion that tends to change the direction of the jet axis of rotation. This precession motion may be due to the multiplicity of the system, which includes the central binary \citep{Sissa2019} and the wide companion (\citealt{Mesa2019}).

This scenario reconciles well with the fact that the environment around R~CrA is rich in diffuse ambient gas and dust that can be swept-up  by powerful jets, as shown by the prominent number of bow shocks, knots, and  outflows detected at different wavelengths (e.g. \citealt{Kumar2011, Anderson1997, Groppi2004, Groppi2007}). Moreover, the shock produced by the wind-carved cavity would be responsible, together with the lower mass wide companion, of the emission in X-ray of hot plasma observed with Chandra and XMM-Newton and analyzed by \citet{Forbrich2006}. The H2 lines may also form in this shock area.

What can our observations tell us about the launching mechanism for the jet of R~CrA? Jets from young stars are usually thought to be launched magneto-centrifugally from disks around them (see e.g. \citealp{Frank2014}). Favored scenarios consider disk wind (see \citealp{Pudritz1983}) or star-disk magnetosphere \citep{Camenzind1990}, the latter in particular in the X-wind scenario \citep{Shu2000}. The disk-wind scenario can itself be separated into two separate schemes, cold-wind and warm-wind \citep{Konigl2000}. As discussed, for example, by \citet{Ferreira2006}, cold winds are generally not considered favorably, because they lead to jets that rotate much faster than usually observed. The main difference between these mechanisms is the launch radius $r_0$, which is related to the specific angular momentum of the jet and to the magnetic leverage $\Lambda$, which is the square of the ratio between the Alfv\`en radius $r_A$, and $r_0$: $\Lambda=(r_A/r_0)^2$. The Alfv\`en radius is where the magnetic energy density is equal to the kinetic energy density. Beyond the Alfv\`en distance, the field lines lag behind the rotation of their footpoints and are coiled into a spiral (see e.g. \citealp{Spruit2010}). We then expect that the radius of the base of the truncated cone representing a jet should be connected to the Alfv\`en radius, though not necessarily be identical, because the jet may have to traverse many Alfv\`en radii before being effectively focused, since its collimation depends not simply on the magnetic lever arm, but also on the poloidal field strength at the disk surface (see \citealp{Frank2014}). The magnetic leverage is in turn related to the ratio between the jet velocity $V_{\rm jet}$\ and the Keplerian $V_{\rm Kep}$\ at $r_0$\ by the relation $V_{\rm jet} = V_{\rm kep}~\sqrt{2 \Lambda -3}$. All these quantities are then related to the launch radius $r_0$. As discussed by \citet{Frank2014}, typical parameters for jets around T~Tau stars are compatible with $r_0$\ of the order of a few stellar radii. The large value of $V_{\rm Kep}$\ at such small separation implies a small value for the magnetic leverage factor ($\sim 5-10$) and slow jet rotation ($\sim 10$~km s$^{-1}$). These are compatible with the very few detections of jet rotation (see e.g. the case of HH212, \citealp{Lee2017}, or the more questioned case of DG~Tau: \citealp{Bacciotti2002, White2014}).

Since R~CrA is a binary with a separation of $a\sim 0.56$~au, which is $\sim 119$~R$_\odot$, it is possible that its jet might be quite peculiar.  The geometrical model discussed in the previous section suggests that the jet is launched close to the primary. However, the presence of a massive enough disk around it is not obvious, because we expect such a disk to be truncated at about a third of the Hill radius, which is at about 20 R$_\odot$ from the star. Since it is likely that the magnetic field of the whole binary system is locked to the binary orbit, the jet orientation would be determined by the orbital phase. On the other hand, the jet velocity is much larger than the escape velocity from the binary. As a consequence, we might expect that the jet describes a helix that may be similar to what is observed. If it exists, such a jet would have properties similar to those observed for other very young stars; in particular, it would be slowly rotating (at most a few tens of km s$^{-1}$),  meaning below the spectral resolution offered by SINFONI. 

As another option, we might perhaps consider the case of a jet that is launched at the inner edge of the circumbinary disk, which is with $r_0$\ of the order of $1.5~a\sim 0.84$~au. Such a jet would have peculiar characteristics. Given the high specific angular momentum of material at this location, we would expect a rapidly rotating jet. The relation between $r_0$\ and the jet rotational velocity $V_{\rm rot}$\ is obtained considering that $r_0=0.05*(2*(l_J/10)/(V_{\rm jet}/115))^2*(M/0.25)^{1/3}$\ \citep{Lee2017}, where $M$\ is the stellar mass and $l_J$\ is the jet specific angular momentum (in au$\times$km s$^{-1}$). Assuming typical values appropriate for the case of R~CrA (jet size $\sim$25~mas, $\sim$4~au; jet velocity $V_{\rm jet}=700$~km s$^{-1}$; stellar mass =$3.0$~M$_\odot$), we obtain $V_{\rm rot}\sim 180$~km s$^{-1}$. This is much larger than typically observed in T~Tau stars, but not incompatible with the value we obtain for the Half Width Half Maximum of the H lines in direction of the jet (and of blobs 1 and 2) that is $\sim 400$~km s$^{-1}$. This is about twice the spectral resolution of SINFONI, and indeed the H lines are clearly much broader than the [Fe~{\sc ii}] or H$_2$ lines (that are not seen in the jet).

Another peculiar property expected for such a jet is the large leverage factor $\Lambda\sim 44$. This would imply $\xi=1/[2~(\Lambda-1)]$=0.01 and $r_A=5.5$~au. $\xi$\ is related to the ratio between the mass-loss rate through the jet and the accretion mass rate, $\xi\sim \dot{M}_{\rm jet}/\dot{M}_{\rm acc}$. Hence a low value of $\xi$\ implies a low value for $\dot{M}_{\rm jet}$, which in turn might explain the low density of the jet of R~CrA. The Alfv\`en radius would be actually similar to the radius of the base of the cone $r_B=5.3$~au derived in our geometrical model. The high value of $\xi$\ is compatible with a well collimated jet (see e.g. \citealp{Garcia2001}). Finally, it should be noted that such a large magnetic leverage factor would be more compatible with a cold-wind jet scenario. 

In summary, the nature of the jet seen in R~CrA is still not clear. Observations are compatible both with the usual warm-disk or X-wind scenarios, but a cold-wind scenario is also possible, related to the peculiar fact that R~CrA is a binary with a separation of a few tens of stellar radii. The two cases might be distinguished by a more accurate estimate of the specific angular momentum of the jet, which might be possible using ALMA (see e.g. \citealp{Lee2017}) for example.

\section{Conclusions} 

Taking advantage of the complementary information provided by optical spectroscopy data (acquired with FEROS) and adaptive optics images in the near infrared obtained with SPHERE and SINFONI, we investigated the extended structure seen around the Herbig Ae/Be star R~CrA.  R~CrA is a very interesting system, not only because it is bright, quite massive, and very young, but also because it is a triple system with a central binary whose separation is of the order of a few tens of the radii of the individual components. This separation might be critical to the survival of circumstellar disks that are considered in the most popular scenarios of magneto-centrifugally launched jets. The vicinity of this star, together with the spatial resolution offered by complementary  instrumentation allowed us to investigate in great detail the extended structures seen around R~CrA. The data reveal a complex overall structure composed by three components: a collimated jet, a wide cavity, and a shock region close to the central system. 

A well  collimated gaseous jet has been detected in the Hydrogen (with SINFONI) and [He~{\sc i}] (with SPHERE) emission lines. It shows a wiggling pattern that is consistent with the period of the binary system. We implemented a geometrical model to reproduce this wiggling pattern, showing that observations may be reproduced by a high velocity ($\sim 770$~km~s$^{-1}$) jet inclined by 3.3$^{\circ}$ toward the observer, which describes a helix on the surface of a cone. The HVC seen in the optical forbidden lines with FEROS is most likely associated with this fast-collimated jet that is flowing inside a cavity carved in the interstellar medium. 

The wide cavity is seen in the continuum emission of the IRDIS and IFS SPHERE data. It shows as a  non-uniform extended structure that extends up to 400~au in the N-E direction, and as wide as $\sim$30$^{\circ}$. The cavity walls are seen in scattered light. The fast-collimated jet appears to be pointing toward the southern side of the cavity, meaning it is not oriented toward the center of the wide cavity, most likely due to a precession motion. [Fe~{\sc ii}] emission is detected along the wall of the cavity. This emission might be attributed to a slower moving wind, most likely a disk wind, that is also producing the LVC seen in the optical forbidden lines. 

The third component is a shock region, close to the central star, where H$_2$ (observed with SINFONI) and possibly the [Fe~{\sc ii}] (observed with FEROS) at 7155~\AA\ are produced at higher density. The velocities of these lines and their FWHM are in between the velocities of the HVC and LVC. 

R~CrA represents a very interesting object, because it allows us to study the structure of jets around Herbig stars, which is not yet well understood. The overall scenario agrees closely with general expectations for magneto-centrifugally launched jets: however, the fact that the star is actually a triple system likely makes the scenario more complex. Given the relevance of this particular object in our understanding of jets from very young stars, more observations would be welcomed to confirm our findings. Namely, a spectroscopic follow-up both with high-resolution spectrographs (in the optical and near-infrared) and with diffraction-limited integral-field spectrographs (e.g. SPHERE, MUSE, and ERIS) may better constrain the regions where the different lines emit, and the kinematic model used to interpret the fast collimated jet. Finally, the launch region of the jet might possibly be established using high-spatial and spectral-resolution observations of the jet with ALMA for example, which may be used to determine the jet rotational velocity.

\begin{acknowledgements}
E.R. is supported by the European Union's Horizon 2020 research and innovation programme under the Marie Sk$\l$odowska-Curie grant agreement No 664931. This work has been supported by the project PRIN INAF 2016 The Cradle of Life - GENESIS-SKA (General Conditions in Early Planetary Systems for the rise of life with SKA) and by the "Progetti Premiali" funding scheme of the Italian Ministry of Education, University, and Research. Programme National de Plan\'etologie (PNP) and the Programme National de Physique Stellaire (PNPS) of CNRS-INSU. This work has also been supported by a grant from the French Labex OSUG\@2020 (Investissements d'avenir - ANR10 LABX56). The project is supported by CNRS, by the Agence Nationale de la Recherche (ANR-14-CE33-0018). This work has made use of the SPHERE Data Centre, jointly operated by OSUG/IPAG (Grenoble), PYTHEAS/LAM CeSAM (Marseille), OCA/Lagrange (Nice), Observatoire de Paris/LESIA (Paris), and Observatoire de Lyon/CRAL. We thank P. Delorme and E. Lagadec (SPHERE Data Centre) for their efficient help during the data reduction process. SPHERE is an instrument designed and built by a consortium consisting of IPAG (Grenoble, France), MPIA (Heidelberg, Germany), LAM (Marseille, France), LESIA (Paris, France), Laboratoire Lagrange (Nice, France), INAF Osservatorio Astronomico di Padova (Italy), Observatoire de Geneve (Switzerland), ETH Zurich (Switzerland), NOVA (Netherlands), ONERA (France) and ASTRON (Netherlands) in collaboration with ESO. SPHERE was funded by ESO, with additional contributions from CNRS (France), MPIA (Germany), INAF (Italy), FINES (Switzerland) and NOVA (Netherlands). SPHERE also received funding from the European Commission Sixth and Seventh Framework Programmes as part of the Optical Infrared Coordination Network for Astronomy (OPTICON) under grant number RII3-Ct-2004-001566 for FP6 (2004-2008), grant number 226604 for FP7 (2009-2012), and grant number 312430 for FP7 (2013-2016). GvdP acknowledges funding from the ANR of France under contract number ANR-16-CE31-0013 (Planet-Forming-Disks). 
\end{acknowledgements}

\bibliographystyle{aa} 
\bibliography{ref_RCrA} %

\end{document}